\documentclass[12pt,preprint]{aastex}
\slugcomment{Revised version 2009 Feb 18 }

\shorttitle{Nearly Periodic X-ray Variability of Blazars}
\shortauthors{Rani, Wiita \& Gupta}

\begin{document}

%% LaTeX will automatically break titles if they run longer than
%% one line. However, you may use \\ to force a line break if
%% you desire.

%%

\title{Nearly Periodic Fluctuations in the Long Term X-ray Light Curves of the Blazars AO 0235$+$164
and 1ES 2321$+$419}

%% Use \author, \affil, and the \and command to format
%% author and affiliation information.
%% Note that \email has replaced the old \authoremail command
%% from AASTeX v4.0. You can use \email to mark an email address
%% anywhere in the paper, not just in the front matter.
%% As in the title, use \\ to force line breaks.

\author{Bindu Rani\altaffilmark{1}, Paul J.\ Wiita\altaffilmark{2,3} and Alok C.\ Gupta\altaffilmark{1}}

\altaffiltext{1}{Aryabhatta Research Institute of Observational Sciences (ARIES),
Manora Peak,  Nainital $-$ 263129, India}
\altaffiltext{2}{School of Natural Sciences, Institute for Advanced Study,
Princeton, NJ 08540}
\altaffiltext{3}{permanent address: Department of Physics and Astronomy, Georgia 
State University, 
Atlanta, GA 30302$-$4106}

\email{bindu@aries.ernet.in, wiita@chara.gsu.edu, acgupta30@gmail.com}%\\
%Phone No. +91-9936683176, Fax No. +91-5942-233439}
%

%% Notice that each of these authors has alternate affiliations, which
%% are identified by the \altaffilmark after each name.  Specify alternate
%% affiliation information with \altaffiltext, with one command per each
%% affiliation.
%
%\altaffiltext{1}{National Astronomical Observatories/Yunnan Observatory, Chinese
%Academy of Sciences, P.O. Box 110, Kunming, Yunnan 650011, China.}
%\altaffiltext{1}{Visiting Astronomer, Cerro Tololo Inter-American Observatory.
%CTIO is operated by AURA, Inc.\ under contract to the National Science
%Foundation.}
%\altaffiltext{2}{Society of Fellows, Harvard University.}
%\altaffiltext{3}{present address: Center for Astrophysics,
%    60 Garden Street, Cambridge, MA 02138}
%\altaffiltext{4}{Visiting Programmer, Space Telescope Science Institute}
%\altaffiltext{5}{Patron, Alonso's Bar and Grill}
%
%% Mark off your abstract in the ``abstract'' environment. In the manuscript
%% style, abstract will output a Received/Accepted line after the
%% title and affiliation information. No date will appear since the author
%% does not have this information. The dates will be filled in by the
%% editorial office after submission.

\begin{abstract}
We have performed a structure function analysis of 
the {\it Rossi X-ray Timing Explorer} All Sky Monitor data 
to search for variability in 24 blazars 
using data trains that each exceed 12 years.
Although 20 of them show nominal periods though this
technique, the great majority of these `periods' are clearly
related to yearly variations arising from the instrument.
Nonetheless, an apparently real periodic component of about 17 days was detected
for the blazar AO 0235$+$164 and it was confirmed by discrete 
correlation function and periodogram analyses.  For 1ES 2321$+$419 a component of variability
with a near periodicity of about 420 days was detected by all of these methods.
We discuss several possible explanations for these nearly
periodic components and conclude that they most likely arise from
the intersections of a shock propagating down a relativistic
jet that possesses a helical structure. 

\end{abstract}

%% Keywords should appear after the \end{abstract} command. The uncommented
%% example has been keyed in ApJ style. See the instructions to authors
%% for the journal to which you are submitting your paper to determine
%% what keyword punctuation is appropriate.

%% Authors who wish to have the most important objects in their paper
%% linked in the electronic edition to a data center may do so in the
%% subject header.  Objects should be in the appropriate "individual"
%% headers (e.g. quasars: individual, stars: individual, etc.) with the
%% additional provision that the total number of headers, including each
%% individual object, not exceed six.  The \objectname{} macro, and its
%% alias \object{}, is used to mark each object.  The macro takes the object
%% name as its primary argument.  This name will appear in the paper
%% and serve as the link's anchor in the electronic edition if the name
%% is recognized by the data centers.  The macro also takes an optional
%% argument in parentheses in cases where the data center identification
%% differs from what is to be printed in the paper.


%\keywords{AGN: blazar: optical: observations - blazars: individual: S5 0716$+$714}
%(\objectname{NGC 6397},
%\object{NGC 6624}, \objectname[M 15]{NGC 7078},
%\object[Cl 1938-341]{Terzan 8})}

%% From the front matter, we move on to the body of the paper.
%% In the first two sections, notice the use of the natbib \citep
%% and \citet commands to identify citations.  The citations are
%% tied to the reference list via symbolic KEYs. The KEY corresponds
%% to the KEY in the \bibitem in the reference list below. We have
%% chosen the first three characters of the first author's name plus
%% the last two numeral of the year of publication as our KEY for
%% each reference.

\section{Introduction}

The characterization of variability timescales can provide information on the sizes 
and locations of the emission regions in active galactic nuclei. Although Doppler
boosted emission from a relativistic jet provides a very reasonable explanation 
for the non-thermal spectra and small-scale radio morphology of the BL Lacertae 
objects and flat spectrum radio quasars that are now usually called blazars 
(e.g., Blandford \& Rees 1978; Urry \& Padovani 1995), the question of just where in such
jets the emission at different wavelengths arises remains somewhat uncertain 
(e.g., Marscher et al.\ 2008).  

While one of the defining characteristics of blazars is extreme variability,
periodic or quasi-periodic contributions to the electromagnetic emission have not been clearly detected, or even claimed to be present, in the vast majority of blazars,
although they certainly have been searched for.  
Probably the best case for such, albeit impermanent, special variations is  S5 0716$+$714, which once showed quasi-periodic variations on the timescale of 1 day, followed by a weaker indication of a variable component of about 7 days, over
the course of an intensive month-long monitoring program.  Quite remarkably,
these fluctuations were present 
simultaneously in an optical and a radio band
(Quirrenbach et al.\ 1991).
On another occasion, quasi-periodicity with a time scale of 4 days seemed to be present 
in the optical band (Heidt \& Wagner 1996). Five major optical outbursts between 
1995 and 2007 seem to occur at intervals of $\sim 3.0 \pm 0.3$ years 
(e.g., Raiteri et al.\ 2003; Foschini et al.\ 2006; Gupta et al.\ 2008a, and references 
therein). Very recently, Gupta, Srivastava \& Wiita (2009) performed a wavelet analysis on the 20 best nights of over 100 high quality optical data sets taken by Montagni et al.\ (2006). They found very high probabilities that S5 0716$+$714 had quasi-periodic components to its intra-night variability on time scales from $\sim$25 to $\sim$73 minutes on several different nights.  

Only one other blazar, OJ 287 (0851$+$203), seems to have shown periodic variations in its light curves over a range of time scales
comparable to that for S5 0716$+$714.  A 15.7 min periodicity 
in 37 GHz radio observations was reported by Valtaoja et al.\ (1985)
for OJ 287. 
In optical bands, a 23 min periodicity was claimed by 
Carrasco, Dultzin-Hacyan, \& Cruz-Gonzalez (1985) and short-lived 32 min periodicity was reported by Carini et al.\ (1992).
Long term optical data on OJ 287 have shown a periodicity of
$\sim$11.7 years; detailed analyses in this case support 
the hypothesis that this
source contains a binary system of supermassive black holes (SMBHs) and 
the major flares arise when the less massive SMBH passes through an
accretion disk surrounding the bigger one
(e.g., Sillanp{\"a}{\"a} et al.\ 1996; Valtonen et al.\ 2008).  
 
A few other blazars may have shown significant periodicity in their flux variations. In the 
blazar PKS 2155$-$304, a quasi-periodicity around 0.7 days seemed to be present in 5 days of observations at UV and optical  wavelengths (Urry et al.\ 1993), and there was a hint that simultaneous x-ray 
observations were well correlated with them (Brinkmann et al.\ 1994). 
One of four $\gtrsim$60 ks x-ray observations of the quasar 3C 273 
by the {\it XMM--Newton} satellite also appears to have a quasi-periodic component with a time scale 
of about 3.3 ks (Espaillat et al.\ 2008).     
A recent analysis of a 91 ks {\it XMM--Newton} observation indicated the presence of a $\sim$ 1 hour  periodicity in the narrow line Seyfert 1 galaxy RE J1034$+$396 
(Gierlinski et al.\ 2008). Using long term (and, unfortunately, very inhomogeneous) optical data on 10 radio-selected blazars, Fan et al.\ (2002) have used the Jurkevich method to claim  detection 
of quasi-periodicity in 9 of them, with putative periods  in the range of 1.4 to 17.9 years. 
   
In \S 2 we discuss the x-ray data 
for 24 blazars stretching over more than 12 years.  
The structure function analyses of these data are given in \S 3.1 and they yield possible quasi-periodicities
for 20 of those objects; however, the great majority of those periods are either $\sim$1 year
or harmonics of an annual period and must be assumed to be an observational artifact.  The structure functions of four objects
showed periods significantly different from a year and we performed additional
analyses of these data (\S \S 3.2 and 3.3)  which strongly 
support the presence of a quasi-period of $\sim$17 days for AO 0235$+$164 and
of $\sim$420 days for 2321$+$419.  These results provide the first good evidences for a nearly periodic component to x-ray
blazar variability  longer than a few hours.  In \S 4 we discuss our results
in terms of several mechanisms that could produce nearly periodic
fluctuations and we
obtain estimates for the central black hole masses in these blazars in the rather
unlikely case that
the observed fluctuations are fundamentally related to orbits of emission 
regions at the inner edges of accretion
disks.  Our conclusions are in \S 5.

\section{Data}

We extracted one day average x-ray fluxes in the 1.5--12 keV energy 
range from the All Sky Monitor (ASM) instrument on board the  {\it Rossi X-ray Timing Explorer}  
satellite\footnote{ASM/RXTE Website: http://xte.mit.edu/ASM\_lc.html}
(RXTE) for the 24 blazars listed in Table 1, which gives their names
and coordinates in the first three columns. 
This data covers the period January 1 1996 through September 1 2008.
These objects are all of the 
blazars in the list of Nieppola, Tornikoski \& Valtaoja (2006) for 
which data is
available on the RXTE web-site; all of them are low- and intermediate-
energy peaked blazars. 
A description of the
ASM and how light curves are obtained from it is given in Levine et al.\ (1996).
In these lengthy data sets,
we found that the source flux counts were given as negative on many days of observations, 
indicating that the source fluxes were then below the detection threshold of the ASM/RXTE. 
Such negative flux counts, or upper limits were omitted in our analysis. We 
first converted the ASM/RXTE fluxes, given in
counts/sec, into a Crab flux unit, using the relation 1 crab = 75 counts/sec; then the flux 
of the source is converted into Janskys, using 
1 crab = 2.4 $\times$ 10$^{-11}$ W m$^{-2}$. 
The x-ray light curves for four of the blazars are presented in Fig.\ 1.

%\begin{table}
%{\bf Table 1. Blazar Structure Function Analysis}

%\vspace*{0.2in}
%\noindent
%\begin{tabular}{lcccc} \hline
%\begin{tabular}{|r|l|c|c|c|} \hline
%Object Name  & $\alpha_{(2000.0)}$ & $\delta_{(2000.0)}$ & Periodicity (days) & F\footnote{Percentage of negative data points} \\\hline

% MOVED TO THE END: RETURN HERE FOR EMULATEAPJ VERSION

%\begin{figure*}
%\plotone{fig1.eps}
%\includegraphics[scale=0.65,angle=-90]{fig1.eps}
%\caption{Partial X-ray light curves from RTXE/ASM for BL Lac, OQ 530, AO 0235$+$164 and
%2321$+$419. Only these blazars show evidence for some non-annual periodicity 
%from a structure function analysis.}
%\end{figure*}

Since the ASM is a survey instrument, and none of these blazars are 
usually among the brightest of x-ray sources, the typical
S/N ($\sim 1.5-3$) is rather poor for each daily data point, though the S/N
is usually substantially higher during the times when the fluxes are
near their peaks. 
Still, the ASM is an unique x-ray instrument, in that it can provide a multi-year light curve of any reasonably strong source.  Our analysis turns out to
yield significant, non-artifactual, periods for only two of the 24 blazars, AO 0235$+$164
and 1ES 2321$+$419 (\S 3).
Recently it has been noticed that a small number of ASM sources have their nominal intensity modulated by the emission of a nearby 
galactic X-ray source (e.g., Kaur et al.\ 2007).   However, the
nearest other source to AO 0235$+$164 observed by 
ASM/RXTE\footnote{http://xte.mit.edu/lcextrct/asmsel.html} is 
separated by $\simeq 4^{\circ}$, which is far enough away that 
it is highly unlikely to contaminate the flux from 0235$+$164.  
In the case of
2321$+$419 the nearest ASM source is separated by over 
$16^{\circ}$ and so  
definitely too far away to cause a contamination problem.

\section{Analyses and Results}

\subsection{Structure Functions}

Ordinary Fourier transform methods are not optimal 
in a search for periodicity in these blazar light curves because the
samplings of these light curves are not exactly uniform. 
Nor can simple periodograms give useful results. Under these circumstances, 
a structure function (SF) analysis is the best way to quantitatively determine any 
time scale of variation  on  unevenly sampled data sets, as these ASM measurements 
have become once we chose to discard the days with ``negative'' fluxes. 

The first order SF is related to the power spectrum density 
(PSD) and discrete correlation function (DCF) and is thus  a powerful tool to search 
for periodicities and time scales in time 
series data (e.g., Simonetti, Cordes \& Heeschen 1985; Gupta et al.\ 2008b and references therein). 
The first order SF for a data set, $a$, having uniformly sampled points is defined as 
\begin{eqnarray}
D^{1}_{a}(k) = {\frac{1}{N^{1}_{a}(k)}} {\sum_{i=1}^N}  w(i)w(i+k){[a(i+k) - a(i)]}^{2},
\end{eqnarray}
where $k$ is the time lag, ${N^{1}_{a}(k)} = \sum w(i)w(i+k)$,
 and the weighting factor $w(i)$
is 1 if a measurement exists for the $i^{th}$ interval, and 0 otherwise.
Since the data in our case is quasi-uniform, we first calucated the differences
squared for all pairs of data points and then averaged the the samples into bins
of one or a few days; measurements are taken to not exist for negative flux values.

Simply summarized, the behavior of the first order SF will at first rise with time lag (after 
a possible plateau arising from noise). Following this rising portion, the SF will then fall into
one of the following classes:
(i) if no plateau exists, any time scale
of variability exceeds the length of the data train;
(ii) if there are one or more plateaus, each one indicates 
a time scale of variability; and
(iii) if that plateau is followed by a dip in the SF, the lag
corresponding to the minimum of that dip, indicates a possible periodic cycle 
(unless such a dip is seen at a lag close 
to the maximum length of the data train, when it is probably an artifact).  

Structure function analyses have been employed for quite some time
in examining the nature of AGN variability.  For example, using long term radio observations of a sample of over 50 radio loud AGN, Hughes, Aller \& Aller (1992) reported that most of them showed some plateau in their SFs; the mean time scale they found for BL Lacs was 1.95 yrs while that for quasars was 2.35 yr. 
A recent extension of this analysis using SFs and other techniques also
examined higher-frequency radio
data and found that small flux density variations were often present on
1 to 2 year time scales but larger outbursts were much rarer; no significant
difference between AGN classes were detected (Hovatta et al.\ 2007).
In a different band, SFs were calculated from a large number of intra-night optical light curves for several blazars (Sagar et al.\ 2004) and for a group of both radio loud and radio quiet quasars as well as blazars (Stalin et al.\ 2005).  Indications of preferred observed time scales of a few hours were found for some objects in each AGN class, and hints of quasi-periods were found for the BL Lac 0851$+$202 and the core-dominated quasars 0846$+$513 and 1216$+$010; however, in none of these cases were more than two dips seen in the SF, so no confident claim of quasi-periodicity could be made based
on those data sets and those SF analyses alone.

The nominal periodicities and approximate standard errors obtained from the SF
analyses for the x-ray light curves of each blazar in our sample are given in the fourth column of Table 1
and the percentage of negative points are listed in the last column.  
We wish to stress that for no blazar have we found a dominant component with a precise
period, but we will henceforth use the words
``period'' and ``periodicity'' to denote the strongest nearly periodic signals seen
in these data sets where the upper-limits are not included in the analysis.
Most of these SF indicated periods are found as averages from at least
two dips and hence standard errors can be obtained, but the period given
for OQ 530 is estimated from a single  dip and so
no errors can be quoted.  The great
majority of the periods in Table 1 are very close to one year.  This is not surprising,
as there has been a previous report of annual, as well as daily, satellite
orbital period (96 minutes) and satellite precession period (53 day) variations
found to be  
imposed on the ASM fluxes of binary X-ray sources 
(Wen et al.\ 2006).  Detection of these periods 
presumably can be attributed to 
windowing effects arising from the satellite and will not be investigated
further.

%\begin{figure*}
%\plotone{fig2.eps}
%\includegraphics[scale=0.67,angle=-90]{fig2.eps}
%\caption{Evidence for a periodic component of $\sim 17$ days
%in the variability of AO 0235$+$164.  The folded light curve is in the upper-left panel and the %structure function (in arbitrary units) is in the lower-left panel.  The discrete correlation %function is in the upper-right panel and
%the spectral power density for a Lomb-Scargle periodogram is
%at the lower-right, with the dotted horizontal lines indicating false alarm probability, $p$, values. 
%}
%\end{figure*}

The SF of the entire RXTE light curve of the blazar AO 0235$+$164 is plotted in the lower-left
panel of Fig.\ 2, binned in 2 day intervals.
This SF shows several significant dips, with the first at about 17 days and the second about 34 days, providing a hint for a periodic component of about 17 days. 
The displayed binning values were chosen for each source so that peaks and dips would be clearest, 
but they remain visible when different bin sizes are used.
The lower-left panel of Fig.\ 3 displays the SF for 2321$+$419, binned in 4 day
intervals; following an almost flat region  consistent with 
noise out to about 50 days the deepest dips are at about 425 and 850 days
but are rather broad. To see whether this hint of a period is genuine, we performed
other analyses on the same data set (\S \S 3.2--3.4).

%\begin{figure*}
%\plotone{fig3.eps}
%\includegraphics[scale=0.67,angle=-90]{fig3.eps}
%\caption{As in Fig.\ 2 for 1ES 2321$+$419, where the periodic component
% is at $\sim 420$ days.}
%\end{figure*}

%\begin{figure*}
%\plotone{fig4.eps}
%\includegraphics[scale=0.67,angle=-90]{fig4.eps}
%\caption{As in Fig.\ 2 for 3C 454.3; this more typical blazar
%shows a very signficant, but irrelevant, annual period.}
%\end{figure*}

In contrast to those two cases, the lower-left
panel of Fig.\ 4 shows the SF for 3C 454.3 (binned over 8 days), for which the 
only  periodicity indicated by multiple clear dips in the SF
is at essentially one year.  This is typical of the SFs of 13 of the 24 cases, where
the only clearly indicated  period is 365 days within a 1$\sigma$ error; in two other cases,
S5 0454$+$844 and 3C 273, the
best value of any period is essentially within 2$\sigma$ of a year and in one case,
BL 1320$+$084, it is roughly 3$\sigma$ away.  All of these nominal periodicities are almost certainly  instrumental `windowing' effects, along with the essentially one-half year periodicities detected for AO 0235$+$164 and OJ 287.  
In four other cases the SF yielded no plausible periodic signal. 
In the last four, most interesting, cases, periods not comparable 
to one year (or one-half year) were found.
Table 2 lists these four  blazars with the most plausible real
periods in column 1 and gives their SF identified  periodicity(-ties) in
column 2.  Their partial light curves are those shown in Fig.\ 1.

%\begin{table}
%{\bf Table 2.  Blazars with Plausible Periodic Components }

%\vspace*{0.1in}
%\noindent
%\begin{tabular}{lccc} \hline
%\begin{tabular}{|r|l|c|c|c|}  \hline
%Object Name & SF (days)&  LSP (days)    & DCF (days)  \\ \hline
%                 &(days)         &Periodicity (days)   &  (days) \\\hline
%AO 0235$+$164  & 17$\pm$1                &   17.8 ($p$ = 0.0294)      & 17              \\ 
%OQ 530        &650, 1302, 910        &960.4 ($p$=0.0848)            & None clear \\ 
%               &                      & (Insignificant peak)                &periodicity  \\ 
%BL Lac        &313$\pm$12      &365.4 ($p$ = 5.82e-7)               & None clear \\ 
%               &                      &                                       &periodicity  \\ 
%2321$+$419      &420               &423.2 ($p$ = 0.00609)          & 430 \\ \hline
%\end{tabular}
%\end{table}

\subsection{Discrete Correlation Functions}

The Discrete Correlation function (DCF) method was first introduced by Edelson \& Krolik (1988) and it was later generalized to provide better error
estimates  (Hufnagel \& Bregman 1992). The DCF is suitable for unevenly sampled data, which is the case in most astronomical observations. In our case, 
as we have discarded the nominally negative counts, the data becomes unevenly sampled. 
Here we give only a brief introduction to the method; for details refer to 
Hovatta et al.\ (2007) and references therein.

The first step is to calculate the unbinned correlation (UDCF) using the given time series through (e.g., Hovatta et al.\ 2007) 
\begin{equation}
UDCF_{ij} = {\frac{(a(i) - \bar{a})(b(j) - \bar{b})}{\sqrt{\sigma_a^2 \sigma_b^2}}},
\end{equation}
where $a(i)$ and $b(j)$ are the individual points in the time series
$a$ and $b$, respectively, $\bar{a}$ and $\bar{b}$ are respectively
the means of the time series and $\sigma_a^2$ and $\sigma_b^2$ are their variances.
The correlation function is binned after calculation of the UDCF. The DCF method does not automatically define a bin size, so several values need
to be tried. If the bin size is too big, useful information is lost, but if the bin size is too small, a spurious correlation can be found. 
For example, we have found that a bin size of 10 days is good
for 2321$+$419  but the minimum bin of 1 day length was best
 for AO 0235$+$164 while 15 days was appropriate for 3C 454.3.  Taking $\tau$ as 
the center of time bin and $n$ is the number of points in each bin, 
the DCF is found from the UDCF via 
\begin{equation}
DCF(\tau) = {\frac{1}{n}} \sum ~UDCF_{ij}(\tau) .
\end{equation}
The error for each bin can be calculated using 
\begin{equation}
\sigma_{\mathrm def}(\tau) = {\frac{1}{n-1}} \Bigl\{ \sum ~\bigl[ UDCF_{ij} - DCF(\tau) \bigr]^2 \Bigr\}^{0.5} .
\end{equation}

A DCF analysis is frequently used for finding the correlation 
and possible lags between 
multi-frequency AGN data where different data trains are used 
in the calculation (e.g., Villata et al.\ 2004; 
Raiteri et al.\ 2003; Hovatta et al.\ 2007 and references therein). 
When the same data train is used, there is obviously
a peak at zero DCF indicating that there is no time lag between the two,
but any other strong peaks in the DCF can indicate a periodicity. 
A disadvantage of this method is that it does not give an exact 
probability that a resulting peak actually represents a periodicity. 
The only way to investigate the internal reliability of the
DCF method is to use simulations; however,
 we have not done so, instead verifying the DCF analysis
by cross-checking the results by the SF and Lomb-Scargle Periodogram
(\S 3.3) methods.  

The results of the DCF analysis for AO 0235$+$164, 1ES 2321$+$419 and 3C 454.3
are shown in the upper-right panels of Figs.\ 2--4, respectively.  
The plotted maximum values of the DCF lags plotted were chosen so 
as to avoid crowded points, but the same features are present in SFs 
extending to the full lengths of the datasets.
The resulting periods for these blazars, along with the negative 
DCF results for OQ 530 and BL Lac, are given in the fourth
column of Table 2.  

\subsection{Lomb-Scargle Periodograms}

The Lomb-Scargle Periodogram (LSP) is another useful technique for searching 
time series for periodic patterns. This method has a good tolerance for missing 
values (e.g., Glynn, Chen \& Mushegian 2006), so it does not require any special 
treatment for gaps in the data and is thus quite suitable for non-uniform data trains.  
Therefore the LSP method is frequently used by astronomers and has
found use in other fields as well (e.g., Glynn et al.\ 2006). 
It also has the advantage of providing a $p$-value which specifies the 
significance of a peak.  The LSP was first introduced by Lomb (1976) 
and later extended by Scargle (1982); somewhat later a more 
practical mathematical formulation was found (Press \& Rybicki 1989).
Here we briefly describe the method and formulae. We used a publicly
available  R language
%(R : A language and environment for statistical computing) 
code for  Lomb-Scargle Periodograms\footnote{http://research.stowers-institute.org/efg/2005/LombScargle}. If $N$ is the total
number of observations,  the LSP is defined at a frequency $\omega_j$
as (Press \& Rybicki 1989; Glynn et al, 2006)  
\begin{displaymath}
P(\omega_j) = {\frac{1}{2 \sigma^2}} \Bigl\{ \frac {(\sum^N_{i=1} [a(t_i)-\bar{a}] \cos[\omega_j(t_i-\tau)])^2}{\sum^N_{i=1} \cos^2[\omega_j(t_i-\tau)]} 
\end{displaymath}
\begin{equation}
+ \frac {(\sum^N_{i=1} [a(t_i)-\bar{a}] \sin[\omega_j(t_i-\tau)])^2} {\sum^N_{i=1} \sin^2[\omega_j(t_i-\tau)]} \Bigr\}.
\end{equation}
Here $j = 1 \dots M$, where $\tau$ is defined by
\begin{equation}
\tan(2 \omega_j \tau) = \frac{\sum_{i=1}^N \sin(2 \omega_j t_i)}{\sum_{i=1}^N \cos(2 \omega_j t_i)},
\end{equation}
and $M$ depends on the number of independent frequencies, $N_0$, through
$M = N_0 \approx -6.363 + 1.193N +0.00098N^2$ (Press et al.\ 2002).

The LSP also provides the ability to test for the presence of more than a single frequency.  We can define a range of frequencies to be tested in the R code for the LSP, and it yields the most significant peak and its significance level.
In searching  for some periodic behavior of a data set, we actually
test the null hypothesis, or false-alarm probability, that the given 
data train is non-periodic at each frequency.  If the probability that the
peak value of the LSP is smaller than $x$, the $p$-value, or probability 
of the  null hypothesis that the observed peak in an LSP was found by chance, is (e.g.. Glynn et al.\ 2006) 
\begin{equation}
p = 1-(1-e^{-x})^M.
\end{equation}
The smaller the $p$-value for a given peak, the higher its significance; 
the maximum limit that can reasonably specified for a $p$-value is 0.05, 
i.e., any  peaks having $p$-values smaller than 0.05 are considered as significant.

Two difficulties usually arise while using such periodograms 
(e.g., Scargle 1982); the first is statistical and the other 
is spectral leakage. The statistical difficulty is mitigated by using large sample
sizes which improves the S/N for possible period detection as this S/N is proportional to 
the number of data points and here we use thousands of points encompassing many
cycles of each putative periodicity.
Spectral leakage, or aliasing, involves leakage of power to some other frequencies
that are actually not present in the data.
A small presence of unevenness in the data spacing substantially reduces aliasing and
astronomical data is typically irregular enough that aliasing is effectively
eliminated.  However, if the sampling is semi-regular (intermediate between randomly
and evenly spaced) significant leakage of periodogram power to
the side-lobes can occur. The usual way to minimize both statistical
and leakage problems is to window or taper the data by smoothing in the spectral domain. 
But the disadvantage of smoothing
is that the spectral values at different frequencies are no longer independent
and hence, the joint statistical properties become more complicated. Since the unevenness
in our x-ray data (after rejection of negative data points) can be best characterized
as random, any leakages of power are expected to be small, and we need not smooth our data.

The results of our LSP analysis, showing the peaks of the
normalized power spectral densities for AO 0235$+$164, 1ES 2321$+$419, and 3C 454.3 
are shown in the bottom-right panels of Figs.\ 2--4, respectively.   
The resulting periods for the first two blazars, along with the
negative results for OQ 530 and BL Lac and their false alarm probablities, $p$, 
are given in the third column of Table 2.

\subsection{\bf Nearly Periodic Variations in Two Blazars}

For all of our sources we have apparent periodic variations present for several
(at least five) cycles. 
We first performed the SF and DCF analyses for the light curves of all the 
sources using unbinned data. Since the data length is large, on these original plots of 
the SF and DCF the points were very crowded. To reduce the data crowding, we binned 
the data in a variety of ways and chose to display the results for lengthy
portions of the data trains and for binning values that provide good clarity in each of the plots. 
The same features remain if the entire non-negative data trains and different bin
sizes are employed. 

To make clearer the scatter about the nearly periodic 
components of the light curves, we have also
plotted folded light curves based on the partial light curves
in Fig.\ 1 for AO 0235$+$164 and 1ES 2321$+$419 and on four cycles
of the quasi-annual variation seen in 3C 454.3.
They are displayed  in the upper-left
panels of Figs.\ 2--4. With a nominal period of 17 days the date of the zero phase on the plot
for AO 0235$+$164 is MJD 50095; with a period of 423 days the plot zero phase for
1ES 2321$+$419 is MJD 50287; using a period of 369 days the zero phase for 3C 454.3
is MJD 50347.  It would be very interesting to see if subsequent observations detect
fluctuations with the same, or similar, periodicities.

Because of the large number of upper-limits in all of the data sets, it must
be noted that our neglecting those values in our analyses is problematic.
For the DCF analysis an alternative, albeit still somewhat arbitrary, approach would be to 
include all such points but to set their values equal to zero.  We did perform such
analyses and in all cases no significant peaks other than at lags of zero were seen and
so any information on periodicity was lost to this form of DCF.
It is to a large extent because of the uncertainties induced by the many upper-limits
and poor individual S/N values that we consider our claimed variations to
be nearly periodic and not overwhelmingly convincing.

\section{Discussion}

\subsection{Properties of the Blazars with Periodic Variability}

The blazar AO 0235$+$164 has a redshift of $z = 0.94$ based on detection of
emission lines (Nilsson et al.\ 1996). Since it was among the first objects to be classified
as a BL Lac (Spinrad \& Smith 1975) and it is a highly variable and rather bright source it has been extensively studied, with over 640 papers mentioning this source.
Its fractional polarization is up to $\sim 40$\% in both the visible and IR bands
(e.g., Impey et al.\ 1982) and it is significantly variable from the
radio to the x-ray bands on timescales ranging from less than an hour to 
many years (e.g., Ghosh \& Soundararajaperumal 1995; Heidt \& Wagner 1996; 
Fan \& Lin 1999; Romero, Cellone \& Combi 2000; Webb et al.\ 2000; Raiteri et al.\ 2001; Padovani et al.\ 2004; Sagar et al.\ 2004; Gupta et al.\ 2008a).  Some of this fast variability is probably due to gravitational microlensing (e.g., Webb et al.\ 2000) as there are 
foreground absorbing systems at $z = 0.524$ and $z = 0.851$ (Burbidge et al.\ 1976).

Raiteri et al.\ (2001) used 25 years of radio and optical data on AO 0235$+$164 
to argue it seemed to have a long
quasi-period of $\sim$5.7 years, but the predicted outburst in 2004 was not detected
(e.g., Raiteri et al.\ 2006).  These authors then suggested a $\sim 8$ year periodicity
might be a better fit to the data.  More recent optical observations provide some measure of support for that suggestion (Gupta et al.\ 2008a).  Our analysis of the archival
RXTE/ASM data provides
the first claim for an x-ray periodicity for this popular blazar.  Using scaling
relations between low-frequency extended radio emission and high frequency beamed
emission (Giovannini et al.\ 2001), Wu et al.\ (2007) have determined a rough value of the Doppler factor for the relativistic
jet of AO 0235$+$164 of $\delta \simeq 10.5$. 

The other blazar that seems show a periodic component to the variability
revealed by the  RXTE/ASM data set is 1ES 2321$+$419
($z = 0.059$; Padovani \& Giommi 1995).  This source is significantly fainter than
AO 0235$+$164 in the optical band and has therefore received much less attention.
Still, it has been studied since its x-ray detection (Elvis et al.\ 1992) in optical (e.g., Falomo \& Kotilainen 1999) and
radio (e.g., Kollgaard et al.\ 1996) bands and a spectral energy distribution is available
(Nieppola et al. 2006).  There have not been any sustained efforts
to look for variability in this blazar in any waveband.  A rather low estimate of $\delta \simeq 1.7$ is
available (Wu et al.\ 2007)

\subsection{Unlikely Explanations for Periodic Variability}

The simplest explanation for such nearly periodic x-ray variability 
in most AGN might be that 
the flux arises from hot spots, spiral shocks or other 
non-axisymmetric phenomena 
related to orbital motions very close to the innermost stable circular orbit around a supermassive black hole (SMBH) 
(e.g., Zhang \& Bao 1991; Chakrabarti \& Wiita 1993; Mangalam \& Wiita 1993).
In the case of AO 0235$+$164 a 17 day period at the inner edge of a disk corresponds to a  SMBH mass of $1.7 \times 10^{9} M_{\odot}$ 
for a non-rotating BH and $1.1 \times 10^{10} M_{\odot}$ for a maximally rotating BH (e.g., Gupta et al.\ 2009).  While the latter mass is quite high, 
the former is a reasonable value,
so it is conceivable that a temporary hot spot in the inner region of an accretion disk is somehow responsible for the observed quasi-periodic variations.
However, for 2321$+$419 a 420 day period yields a SMBH mass of $7.5 \times 10^{10} M_{\odot}$ for a non-rotating BH and $4.8 \times 10^{11} M_{\odot}$ for a maximally rotating BH. 
In this case, in order to reduce the SMBH mass to a quite high, but conceivable, value of $3 \times 10^9 M_{\odot}$, in either case of
BH spin the dominant hot spot would need to be located at a distance of $\sim 51 GM/c^2$ 
from the SMBH, which seems to be rather far away for a hot spot that contributes significant flux.  Another reason to 
discount this hot spot scenario for both blazars is that blazar disks are almost certainly close to face-on but large hot spot amplification arises from near-field gravitational 
lensing and that is strong only if the observer's line-of-sight is close to the disk plane (e.g., Bao, Wiita \& Hadrava 1996).

A somewhat related possibility is that we are seeing the interaction between a second black hole and the disk surrounding the primary one, 
as seems to be the case for OJ 287 (e.g.\ Valtonen et al.\ 2008).
However, such an orbital cycle should probably yield a more precise 
and long-lived periodicity than we have found for the x-ray emission of both AO 0235$+$164 and 2321$+$419, so we believe this hypothesis is quite unlikely.  In addition, for AO 0235$+$164 the rather
short period strongly disfavors the binary black hole hypothesis.

The microlensing hypothesis does not appear to be able to produce
a quasi-periodic component to the variability, so even if it does play a
role in producing some variability in the observed flux of
AO 0235$+$164, it probably
is irrelevant to the fluctuations of interest here. 

\subsection{More Likely Explanations}

As the preponderance of other evidence has the x-rays seen
from blazars, particularly in active phases, emerge 
from their jets, and not their putative accretion disks or coronae, 
it makes sense to examine how such 
quasi-periods could be related to jet structures.  
Turbulence behind a shock propagating 
down a jet (e.g., Marscher,
Gear \& Travis 1992) is a very logical, but not yet carefully treated,
 way to produce variability.  For such turbulent flows the dominant eddies' 
turnover times should yield short-lived, quasi-periodic, but probably 
modest, fluctuations in emissivity.  Regions at different distances
behind the shock will emit preferentially at different wavelengths.  
But because Doppler boosting can greatly
amplify (roughly by a factor of $\delta^2$ to $\delta^3$, e.g., 
Blandford \& Rees 1978) even weak intrinsic flux 
variations produced by small
changes in the magnetic field strength or relativistic electron density, 
can be raised to the 
level at which they can be detected (e.g., Qian et al.\ 1991).  
This same Doppler boosting reduces the time-scale at which these
fluctuations are observed by a factor of $\delta$
compared to the time-scale they
possess in the emission frame.  Although it is difficult to quantify these 
effects precisely, this mechanism does seem to provide an 
excellent way to understand the optical 
intra-night variability with quasi-periods of tens of minutes 
that are only occasionally 
seen and that have timescales that vary from night to night in
the blazar S5 0716$+$714 (Gupta et al.\ 2009).

The same turbulence in shocked-jet scenario could  be playing out
in a blazar such as
AO 0235$+$164, where the fairly high value of $\delta \approx 10$
easily allows for modest fluctuations to become easily visible; however,
the observed period of $\sim 17$ days converts into an eddy turnover
time of $\sim 87$ days in
the rest frame for such a Doppler factor.  This would require a much larger,
but still reasonably sized, eddy to be involved.

For 2321$+$419 this turbulent jet explanation is somewhat less likely to work if
the Doppler factor is only $\sim 1.7$, as it would only produce
amplifications of 5 or less.  Moreover, the nominal rest-frame 
eddy turnover time would
be $\sim 675$ days, which implies quite a substantial sized eddy and
that the x-ray variations were arising at distances $> 1$ pc
from the nucleus.  

It is quite likely that blazar jets will possess some 
essentially helical structure, such as can be easily induced by 
magnetohydrodynamical instabilities in a magnetized jet 
(e.g., Hardee \& Rosen 1999) or through precession.  Indeed,
in the few cases where the innermost portions of
radio jets can be resolved transversely
using VLBI,  edge-brightened
and non-axisymmetric structures are seen 
(e.g., M87, Ly, Walker \& Junor 2007; Cen A, Bach et al.\ 2008; 
Mkn 501, Piner et al.\ 2009).

A relativistic shock propagating down such a perturbed jet will 
induce significantly increased emission at the locations
where the shock intersects a region of enhanced magnetic field and/or electron density corresponding to
such a non-axisymmetric structure.
Thanks to the extreme sensitivity of Doppler boosting to
viewing angle, very substantial  
changes in the amplitude (and polarization) of radio and optical
jet emission will be seen by an observer at fixed angle
to the jet axis as the most strongly emitting region effectively
swings past the observer (e.g., Camenzind \& Krockenberger 1992; 
Gopal-Krishna \& Wiita 1992).   
There is no reason why the
intersection of a relativistic shock with a quasi-helical
perturbation would not perform similar
feats for the x-ray emission, even though these high energy photons 
are unlikely to emerge from
exactly the same jet regions as the optical and radio photons (e.g.,
Marscher et al.\ 2008) and should have somewhat different
temporal dependences.  

Because of the apparently large Doppler factor of the jet of
AO 0235$+$164,  the observed substantial and 
nearly periodic components in its x-ray light curve can
be naturally  attributed to
the intersections of a relativistic shock with successive 
twists of a non-axisymmetric jet structure.
The apparently modest Doppler factor of the jet in 2321$+$419
makes this explanation less immediately attractive; however,
all other hypotheses work even less well for this source if the
estimated low Doppler factor is correct.  So the intersection
of a shock with a non-axisymmetric jet structure also seems 
to be the most plausible explanation for the behavior of 
this blazar.

\section{Conclusions}

We searched the RXTE/ASM light curves extending over 12 years 
of 24 blazars for possible periodic
variations using structure functions.  Many of them showed apparent periods,
but the majority of these were close to one year and presumably not real.
The four blazars that showed indications of non-artifactual periods
were examined further using discrete correlation functions and 
Lomb-Scargle periodograms.

Two blazars showed nearly common periodic components to their x-ray variability 
through all three methods and had low ($<0.03$) false alarm probabilities
according to the LSP method: AO 0235$+$164 shows an observed 
period of $\sim 17$ days while 1ES 2321+419 has one of $\sim 420$ days.

It is quite unlikely that these nearly periodic fluctuations are
caused by orbiting hot spots on or above accretion disks or by
a companion black hole crashing through an accretion disk on each
orbit.  It is even
less likely that these fluctuations  are produced by microlensing.  

Turbulence
behind a shock moving through a relativistic jet may provide an adequate
explanation of our results
if the variations are dominated by large-scale eddies moving
into and out from our line of sight.  Still, the most attractive
hypothesis to explain these variations appears to be the intersection
of a shock with an essentially helical structure wrapping around
the relativistic jet. In this case,  x-ray polarimetry variations should be
correlated with the flux changes (e.g., Gopal-Krishna \& Wiita 1992)
and might eventually provide a way to 
distinguish between these different possible explanations.

\acknowledgments

We thank the referee for several suggestions that improved the presentation of
the results.
PJW's work is supported in part by a subcontract to GSU from NSF grant AST 05-07529 
to the University of Washington. This research has made use of data obtained through 
the High Energy Astrophysics Science Archive Research Center Online Service, provided 
by NASA's Goddard Space Flight Center (GSFC). ASM results were provided by the ASM and 
RXTE teams at MIT and at the RXTE SOF and GOF at NASA's GSFC. This research has made use of the NASA/IPAC Extragalactic Database (NED) which is operated by the Jet Propulsion Laboratory, California Institute of Technology, under contract with NASA.

\clearpage
\begin{table}
{\bf Table 1. Blazar Structure Function Analysis}

\vspace*{0.2in}
\noindent
\begin{tabular}{lcccc} \hline
%\begin{tabular}{|r|l|c|c|c|} \hline
Object Name  & $\alpha_{(2000.0)}$ & $\delta_{(2000.0)}$ & Periodicity (days) & F$^a$  \\\hline
1ES 0145+138 &01h 48m 29.7s & +14$^{\circ}$ 02$^{\prime}$ 18$^{\prime \prime}$ & 365$\pm$7 & 42.6 \\ 
3C 66A       &02h 22m 39.6s & +43$^{\circ}$ 02$^{\prime}$ 08$^{\prime \prime}$ & 369$\pm$16  & 38.8 \\ 
AO 0235+164  &02h 38m 38.8s & +16$^{\circ}$ 36$^{\prime}$ 59$^{\prime \prime}$ & 17$\pm$1, 162$\pm$4, 275$\pm$20 & 41.8 \\ 
S5 0454+844  &05h 08m 42.5s & +84$^{\circ}$ 32$^{\prime}$ 05$^{\prime \prime}$ & 336$\pm$14 & 42.3 \\ 
S5 0716+714  &07h 21m 53.3s & +71$^{\circ}$ 20$^{\prime}$ 36$^{\prime \prime}$ & 347$\pm$18 & 37.6 \\ 
PKS 0735+178 &07h 38m 07.4s & +17$^{\circ}$ 42$^{\prime}$ 19$^{\prime \prime}$ & 347$\pm$31 & 41.8 \\ 
PKS 0829+046 &08h 31m 48.9s & +04$^{\circ}$ 29$^{\prime}$ 39$^{\prime \prime}$ & 364$\pm$5 & 39.9 \\ 
OJ 287       &08h 54m 48.8s & +20$^{\circ}$ 06$^{\prime}$ 30$^{\prime \prime}$ & 148$\pm$19, 337$\pm$26 &43.6  \\ 
S4 0954+658  &09h 58m 47.2s & +65$^{\circ}$ 33$^{\prime}$ 54$^{\prime \prime}$ & None detected & 42.6 \\ 
BL 1147+245  &11h 50m 19.2s & +24$^{\circ}$ 17$^{\prime}$ 54$^{\prime \prime}$ & 359$\pm$21 & 38.6 \\ 
1ES 1212+078 &12h 15m 10.9s & +07$^{\circ}$ 32$^{\prime}$ 03$^{\prime \prime}$ & 362$\pm$26 & 39.4 \\ 
ON 231       &12h 21m 31.7s & +28$^{\circ}$ 13$^{\prime}$ 58$^{\prime \prime}$ & 367$\pm$7 & 46.5 \\ 
3C 273       &12h 29m 06.7s & +02$^{\circ}$ 03$^{\prime}$ 09$^{\prime \prime}$ & 391$\pm$15 & 17.4 \\ 
3C 279       &12h 56m 11.2s & $-$05$^{\circ}$ 47$^{\prime}$ 22$^{\prime \prime}$ & 361$\pm$26 & 43.1 \\ 
BL 1320+084  &13h 22m 54.9s & +08$^{\circ}$ 10$^{\prime}$ 10$^{\prime \prime}$ & 337$\pm$8 & 43.3 \\ 
OQ 530       &14h 19m 46.6s & +54$^{\circ}$ 23$^{\prime}$ 14$^{\prime \prime}$ & 650, 910, 1300 & 41.7 \\ 
PG 1553+11   &15h 55m 43.1s & +11$^{\circ}$ 11$^{\prime}$ 24$^{\prime \prime}$ & 363$\pm$18 & 31.5 \\ 
BL 1722+119  &17h 25m 05.5s & +11$^{\circ}$ 52$^{\prime}$ 16$^{\prime \prime}$ & 361$\pm$26 & 34.8 \\ 
3C 371       &18h 06m 50.7s & +69$^{\circ}$ 49$^{\prime}$ 28$^{\prime \prime}$ & None detected & 38.0 \\ 
S5 2007+77   &20h 05m 31.1s & +77$^{\circ}$ 52$^{\prime}$ 43$^{\prime \prime}$ & None detected & 39.3 \\ 
BL Lac       &22h 02m 43.3s & +42$^{\circ}$ 16$^{\prime}$ 39$^{\prime \prime}$ & 313$\pm$12 & 36.0 \\ 
3C 454.3     &22h 53m 57.7s & +16$^{\circ}$ 08$^{\prime}$ 54$^{\prime \prime}$ & 361$\pm$3 & 37.0 \\ 
1ES 2321+419 &23h 23m 54.1s & +42$^{\circ}$ 11$^{\prime}$ 19$^{\prime \prime}$ & 425$\pm$10 &33.3 \\ 
1ES 2344+514 &23h 47m 04.8s & +51$^{\circ}$ 42$^{\prime}$ 18$^{\prime \prime}$ & None detected & 33.3  \\ \hline
\end{tabular}
$^a$Percentage of negative data points
\end{table}
\clearpage

\begin{table}
{\bf Table 2.  Blazars with Plausible Periodic Components }

\vspace*{0.1in}
\noindent
\begin{tabular}{lccc} \hline
%\begin{tabular}{|r|l|c|c|c|}  \hline
Object Name & SF (days)&  LSP (days)    & DCF (days)  \\ \hline
%                 &(days)         &Periodicity (days)   &  (days) \\\hline
AO 0235$+$164  & 17$\pm$1                &   17.7 ($p$ = 0.0294)      & 17              \\ 
OQ 530        &650, 1302, 910        &960.4 ($p$=0.0848)            & None clear \\ 
%               &                      & (Insignificant peak)                &periodicity  \\ 
BL Lac        &313$\pm$12      &365.4 ($p$ = 5.82e-7)               & None clear \\ 
%               &                      &                                       &periodicity  \\ 
2321$+$419      &425$\pm$10               &423.2 ($p$ = 0.00609)          & 430 \\ \hline
\end{tabular}
\end{table} 

\clearpage
\begin{figure*}
\plotone{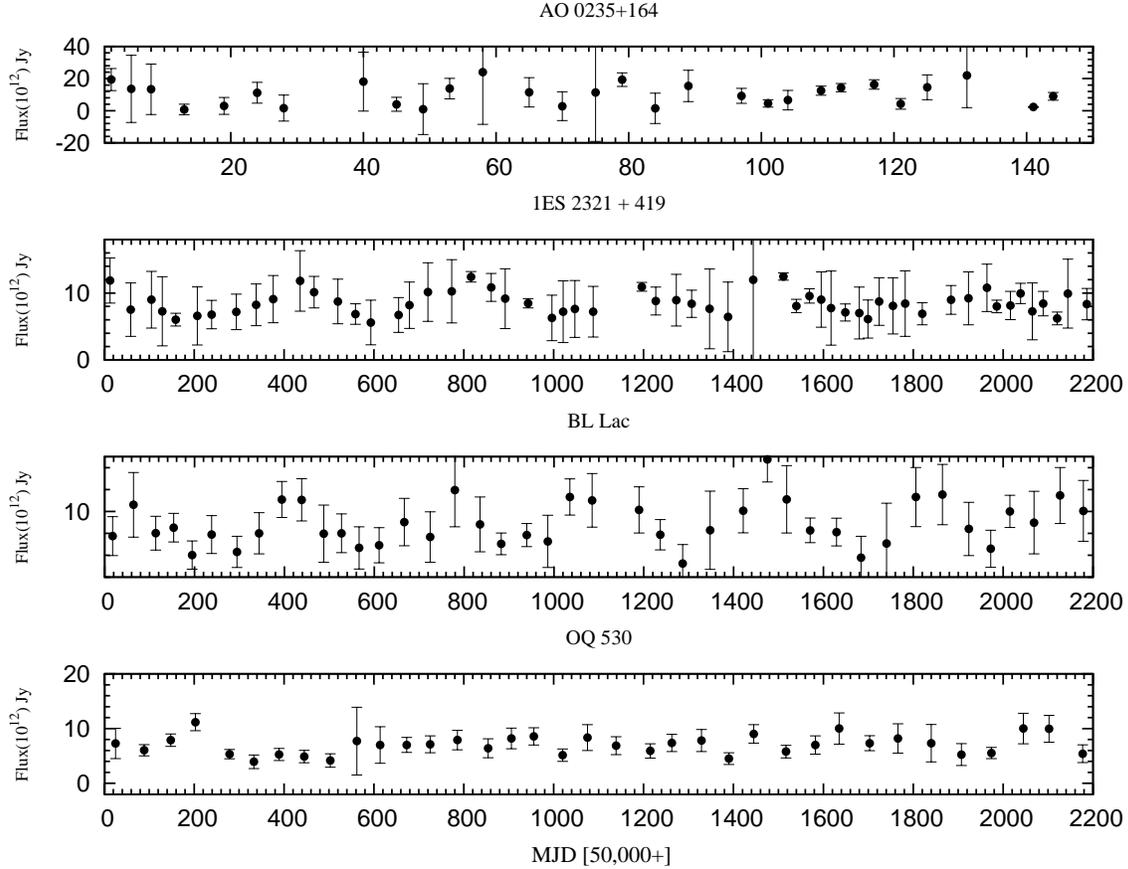}
\caption{Partial X-ray light curves from RTXE/ASM for AO 0235$+$164, 2321$+$419, BL Lac and OQ 530. 
Only these blazars show evidence for some non-annual periodicity from a structure function analysis. 
Variability in AO 0235$+$164 and 1ES 2321$+$419 was detected by all the three analysis methods whereas
the variability in BL Lac and OQ 530 was identified by SF and LSP methods only.}
\end{figure*}

\clearpage
%\noindent
\begin{figure*}
%\plotone{fig2.eps}
\includegraphics[scale=0.67,angle=-90]{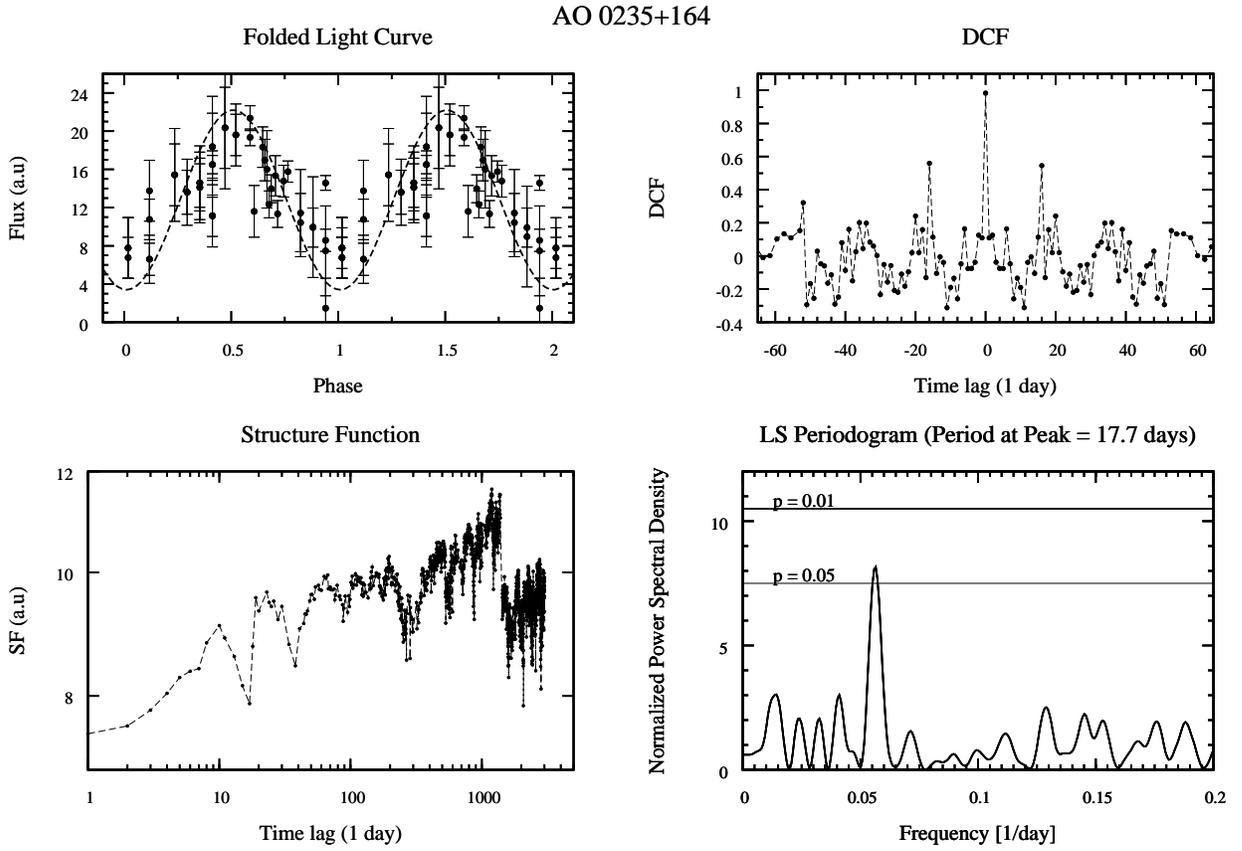}
\caption{Evidence for a periodic component of $\sim 17$ days
in the variability of AO 0235$+$164.  The folded partial light curve is in the upper-left panel and the structure function (in arbitrary units) is in the lower-left panel.  The discrete correlation function is in the upper-right panel and
the spectral power density for a Lomb-Scargle periodogram is
at the lower-right, with the  horizontal lines indicating false alarm probability, $p$, values. 
}
\end{figure*}

\clearpage
\begin{figure*}
%\plotone{fig3.eps}
\includegraphics[scale=0.67,angle=-90]{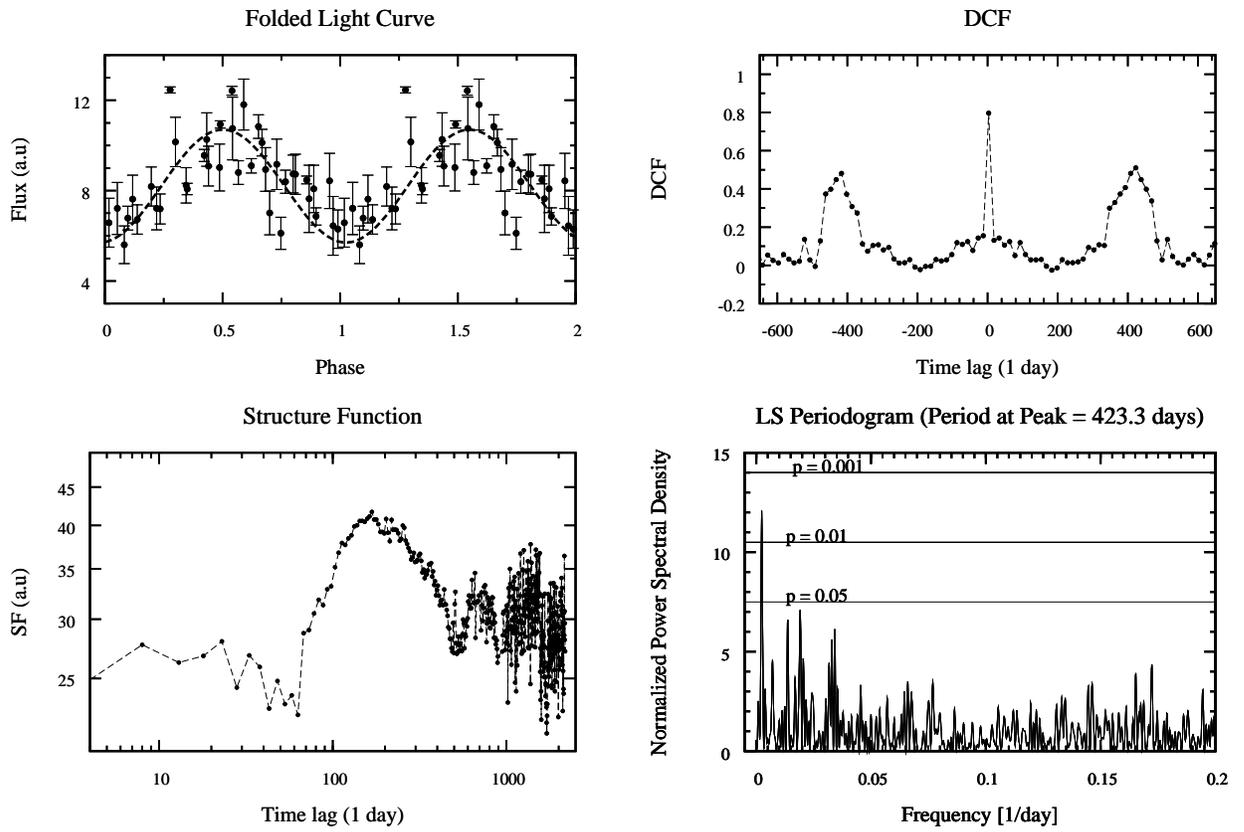}
\caption{As in Fig.\ 2 for 1ES 2321$+$419, where the periodic component
 is at $\sim 423$ days.}
\end{figure*}

\clearpage
\begin{figure*}
%\plotone{fig4.eps}
\includegraphics[scale=0.67,angle=-90]{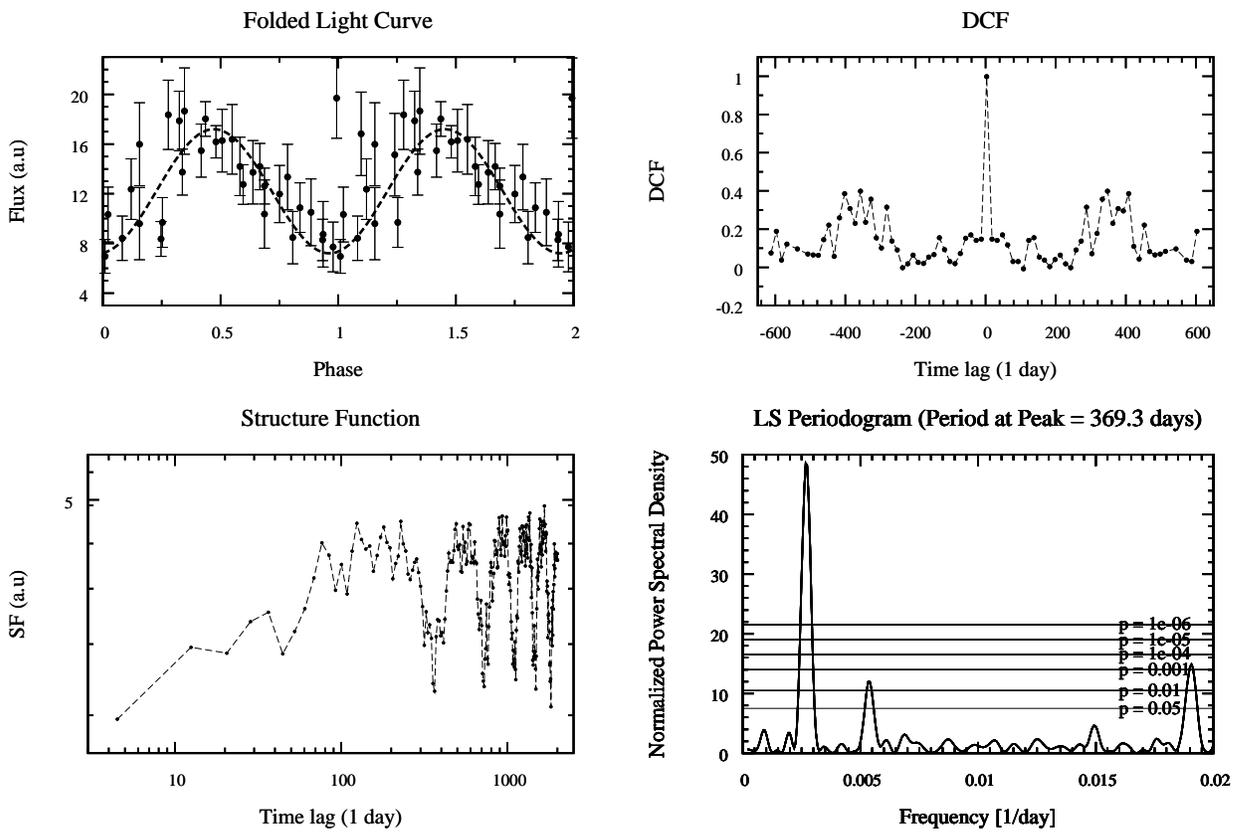}
\caption{As in Fig.\ 2 for 3C 454.3; this more typical blazar
shows a very signficant, but irrelevant, nearly annual period.}
\end{figure*}

\end{document}